\documentstyle[12pt,draft]{article}
\input epsf

\newcommand{\resection}[1]{\setcounter{equation}{0}\section{#1}}

\def\t{\theta}

\def\e{\epsilon}

\def\s {\sigma}

\def\be{\begin{equation}}
\def\ee{\end{equation}}
\def\bea{\begin{eqnarray}}
\def\eea{\end{eqnarray}}
\def\beano{\begin{eqnarray*}}
\def\eeano{\end{eqnarray*}}
\def\bd{\begin{displaystyle}}
\def\ed{\end{displaystyle}}
\def\ba{\begin{array}}
\def\ea{\end{array}}

\def\nn{\nonumber}
\def\nonu{\nonumber\\}
\def\nv{\vec {\bf n}}
\def\kv{\vec {\bf k}}
\def\nls{nl$\s\;$}

\def\x{|_{x=0}}

\def\pd{\partial}

\def\NPB{{ Nucl. Phys.} \bf B}
\def\PLB{{ Phys. Lett.}  \bf B}

\def\PRD{{ Phys. Rev.} \bf D}

\def\IJMPA{{ Int. Jour. Mod. Phys.} \bf A}

\def\CMP{ Comm. Math. Phys. \bf}

\def\AnPh{Ann. Phys. }

\begin{document}
\oddsidemargin 5mm
\setcounter{page}{0}
\newpage     
\setcounter{page}{0}
\begin{titlepage}
\begin{flushright}
ICTP/IR/98/19\\
SISSA/EP/98/101
\end{flushright}
\vspace{0.5cm}
\begin{center}
{\large {\bf Quantum Integrability of Certain Boundary Conditions}}
\footnote{Work done under 
partial support of the EC TMR Programme {\em Integrability, 
non--perturbative effects and symmetry in Quantum Field Theories}, grant
FMRX-CT96-0012.} \\
\vspace{1.5cm}
{\bf 
M. Moriconi} \footnote{\tt{email:moriconi@ictp.trieste.it}}\\
{\em The Abdus Salam International Centre for Theoretical Physics}\\
{\em Strada Costiera 11, 34100 Trieste, Italy}\\
\vspace{0.8cm}
{\bf 
A. De Martino} \footnote{\tt{email:demarti@sissa.it}}\\
{\em International School for Advanced Studies, Via Beirut 2-4, 
34014 Trieste, Italy} \\
{\em Istituto Nazionale di Fisica Nucleare, Sezione di Trieste}\\
\end{center}
\renewcommand{\thefootnote}{\arabic{footnote}}
\vspace{6mm}

\begin{abstract}
\noindent
We study the quantum integrability of the $O(N)$
Nonlinear $\s$ (nl$\s$) model and the $O(N)$ Gross-Neveu (GN) model on the 
half-line. We show that the \nls model 
is integrable with Neumann, Dirichlet and
a mixed boundary condition and that the GN model
is integrable if $\psi_+^a\x=\pm\psi_-^a\x$. We also comment on the 
boundary condition found by Corrigan and Sheng for the $O(3)$ \nls model.

\vspace{3cm}

\end{abstract}
\vspace{5mm}
\end{titlepage}

\newpage
\setcounter{footnote}{0}
\renewcommand{\thefootnote}{\arabic{footnote}}

\resection{Introduction}

The purpose of this letter is to investigate the quantum integrability of
certain boundary conditions of two theories defined on the half-line 
\cite{gzam,FrKob}:
the $O(N)$ nonlinear sigma model (nl$\s$) and the 
$O(N)$ Gross-Neveu (GN) model. These two models
have some very similar properties, such as 
asymptotic freedom and dynamical mass
generation, but are also quite different, the former being bosonic 
and no bound states,
and the latter fermionic and with a very rich spectrum
of bound states, for example. 
Their bulk version has been established to be integrable
long ago, at the classical \cite{Pohlmeyer,NevPap} and at the 
quantum level \cite{Polyakov,Witten}. 
The study of these models on the
half-line is hindered more difficult, because many of the 
techniques available
on the full-line, such as the Lax pair, 
cannot be easily extended to the half-line.
The structure of this letter is as follows.
In the next section we briefly
review the two models and exhibit a bulk conserved charge of spin 3
for each model, in section 3 we show that the Neumann, Dirichlet and
the ``mixed''  boundary condition
preserve integrability for the \nls  model, and discuss the 
condition found by Corrigan and Sheng in \cite{CorrSheng} for the
$O(3)$ \nls model; in section 4 we show that 
the GN model on the half-line is integrable if
$\psi_+^a\x=\pm \psi_-^a\x$, where 
$\psi_{\pm}^a$ are the chiral components
of the Majorana fermions. In the final
section we present our conclusions and possible extensions 
of this work.

\resection{The Models}

In this section we briefly review the main properties of the \nls
model and of the GN model. We also discuss the conserved currents
of spin 4 that are going to be used later.

\subsection{The $O(N)$ Nonlinear $\s$ Model}

The $O(N)$ nonlinear $\sigma$ (\nls) model is defined by the 
following Lagrangian
\be
{\cal L}_{nl\s}=\frac{1}{2 g_0} \pd_\mu \nv \cdot \pd^\mu \nv
\ , \ \label{nlsm}
\ee
where $\nv$ is a vector in $N$-dimensional space, subject to the constraint
$\nv \cdot \nv=1$ \footnote{Our conventions throughout this paper are: The 
Minkowski metric is $\eta_{\mu \nu}=$ diag(-1,1), the gamma matrices are
$\gamma^0=i\sigma_2$ and $\gamma^1=\sigma_1$, where $\sigma_i$ are the Pauli
matrices. The light-cone variables are
$x_{\pm}=(x_0 \pm x_1)/2$.}. 
We can introduce a Lagrange multiplier $\omega$ that takes care of the
constraint $\nv \cdot \nv=1$, the Lagrangian being modified to
\be
{\cal L}'_{nl\s}=\frac{1}{2 g_0} \pd_\mu \nv \cdot \pd^\mu \nv+
\omega (\nv \cdot \nv -1) \ . \
\ee
By using this Lagrangian and the constraint on the length of $\nv$ it is easy
to show that the classical equation of motion in light-cone coordinates is
\be
\pd_+\pd_-\nv+\nv(\pd_-\nv \cdot \pd_+ \nv)=0 \ . \ \label{eomnls}
\ee
At the classical level this model is conformally
invariant, which implies the vanishing of the off-diagonal
components of the energy-momentum tensor
\be
T_{+-}=T_{-+}=0 \ .
\ee
The non-vanishing components are
\be
T_{++}=\pd_+ \nv \cdot \pd_+ \nv \qquad {\rm and} \qquad
T_{--}=\pd_- \nv \cdot \pd_- \nv \ . \ 
\ee
Notice that the conservation law 
\be
\pd_{\pm}T_{\mp \mp}=0 \label{confinv}
\ee
implies $\pd_{\pm}(T_{\mp \mp})^n=0$ for any integer $n$.

At the quantum level there is dynamical mass generation.
This means that this model has an anomaly and so,
conserved charges in the classical theory have to be corrected. It is not
clear in principle that this model will be still integrable.
Nonetheless Polyakov proved the quantum integrability of the \nls model 
in \cite{Polyakov} (see also \cite{ZZAnn}).

In \cite{GoldWitt} Goldschmidt and Witten 
have analyzed conserved charges of some two-dimen\-sional models
in a similar way as Polyakov did for the \nls model, and showed their 
quantum integrability. 
Their argument goes as follows.
Since the theory is anomalous, 
the right hand side of $\pd_{\pm}(T_{\mp \mp})^2=0$
is not zero 
anymore, and we have to include, in principle, all possible operators of
dimension $5$ and Lorentz weight $\mp 3$. 
We should then list these operators and 
check which ones can be written as total derivatives. In the case of the \nls
model they showed that $\pd_{\pm}(T_{\mp \mp})^2=0$ 
can only pick up anomalous
contributions that can be written as total derivatives, namely $\pd_{\pm}$ of
something. So the classical conservation 
law is inherited to the quantum level. 

If we take the spin $3$ conservation law, 
$\pd_{\pm}(T_{ \mp \mp})^2=0$, our
previous discussion shows that at after quantization it becomes
\be
\pd_{+}(T_{--})^2=c_1 \;\pd_+ (\pd_{-}^2\nv \cdot \pd_{-}^2\nv)+
c_2\;\pd_- (\pd_+\nv \cdot \pd_-\nv \; \pd_-\nv \cdot \pd_-\nv)+
c_3\;\pd_-(\pd_-^3\nv \cdot \pd_+\nv) \ , \ \label{chargenls}
\ee
where the $c_i$ are constants. Of course we have a similar expression for
$\pd_-(T_{++})^2$, taking $+ \leftrightarrow -$,
with the same coefficients $c_i$. This result implies
the existence of two nontrivial charges in the \nls model at the quantum level,
and therefore its integrability \cite{Parke}.

The boundary version of the \nls model was first considered by Ghoshal in 
\cite{Ghoshal}, where it is also conjectured the integrability of the 
Neumann, ($\pd_1 \nv\x=0$) and Dirichlet ($\pd_0 \nv \x=0$), boundary 
conditions. 
In that paper Ghoshal solved the boundary 
Yang-Baxter equation consistent with this choice (and this
is the main argument for its integrability!). 
The classical integrability of the Neumann condition for the $O(N)$ \nls model
was established by means of a generalization of the Lax pair to the half-line
by Corrigan and Sheng in \cite{CorrSheng}.

\subsection{The $O(N)$ Gross-Neveu Model}

The Gross-Neveu model \cite{GrNev} is a fermionic theory with quartic Fermi coupling defined
by the following Lagrangian
\be
{\cal L}_{gn}=i\bar{\psi}\not\!\pd \psi
+\frac{g^2}{4}(\bar{\psi}\psi)^2 \ , \
\label{gn}
\ee
where $\psi$ is a $N$ component Majorana spinor in the fundamental 
representation of $O(N)$, with components $\psi^a$, $a$ from $1$ to $N$. 
The chiral components of the $\psi^a$ are 
$(\psi_+^a,\psi_-^a)$. In light-cone coordinates the
GN model Lagrangian becomes
\be
{\cal L}_{gn}=-\psi_+^a i\pd_- \psi_+^a -\psi_-^a i\pd_+ \psi_-^a +
g^2(\psi_+^a\psi_-^a)^2 \ . \
\ee
Notice that $\pd_{\pm}\rightarrow \exp(\pm \t)
\pd_{\pm}$ and $\psi_\pm^a \rightarrow \exp(\pm \t/2)\psi_\pm^a$, 
under a Lorentz transformation \footnote{$\t$ is the rapidity variable 
parameterizing the Lorentz transformation.}. This means
that $\psi_\pm$ has Lorentz weight $\pm 1/2$, and $\pd_\pm$ has
Lorentz weight $\pm 1$.
The equations of motion are
\be
i\pd_\mp \psi_\pm^a=\pm g^2 \psi_\mp^a (\psi_+^b\psi_-^b) \ . \ \label{eomgn}
\ee

The classical integrability of this model was established 
by Neveu and Papanicolaou in \cite{NevPap}.
The quantum integrability of the GN model 
was established in \cite{ZamZam}, where it was proved, in the
large $N$ limit, that there is no
particle production. The construction of quantum conserved charges for the
GN model was done in
an analogous way to Polyakov's 
construction for the \nls model, in \cite{Witten}.

Following Witten
\cite{Witten}, we start by looking at the classical conservation laws
due to the conformal invariance of \ref{gn}. The diagonal components of
the energy-momentum tensor
are $T_{\pm \pm}=\psi^a_{\pm}\pd_{\pm} \psi^a_{\pm}$, the off-diagonal
components, $T_{+-}$ and $T_{-+}$ being zero. 
Let us consider the spin 3 conservation law, $\pd_-(T_{++})^2=0$.
The left hand side of this equation has dimension
5 and Lorentz weight 3. This implies that
the possible anomalies have to be either linear in $\pd_-$ and zeroth
order in $\psi_-^a$ or zeroth order in $\pd_-$ and quadratic in $\psi_-^a$.
The operators of the former type can be converted into operators of the latter
type by using the equations of motion.  
Analyzing all local operators with the required properties,
Witten showed in \cite{Witten} 
that these terms can be written
as total derivatives. We list these terms in the appendix.
This means that anomalies destroy conformal invariance but 
do not destroy the conservation law, 
and the GN model is integrable at the quantum level.

\resection{Integrable Boundary Conditions \\ for GN and \nls  Models}

When considering the boundary version of an integrable field theory not all
charges will still be conserved. Therefore one
should investigate whether some combination of the bulk
charges can be preserved
after the introduction of a boundary. If we have some spin $s$ 
conservation law of the form
\be
\pd_-J_+^{(s+1)}=\pd_+ R_-^{(s-1)}\qquad {\rm and} \qquad 
\pd_+J_-^{(s+1)}=\pd_- R_+^{(s-1)} \ , \
\ee
then we know that
\be
Q_+=\int_{-\infty}^{+\infty} dx_1 \; (J_+^{(s+1)}- R_-^{(s-1)} ) 
\qquad {\rm and} \qquad
Q_-=\int_{-\infty}^{+\infty} dx_1 \; (J_-^{(s+1)}- R_+^{(s-1)} )
\ee
are conserved, $\pd_0 Q_\pm =0$. In proving that these charges are conserved we
have to use the fact that we can discard surface terms. When we restrict our
model to the half-line we can not do that with one of the surface terms.
On the other hand, if the following condition \cite{gzam} is satisfied
\be
J_{-}^{(s+1)} -J_{+}^{(s+1)} +
R_{-}^{(s-1)} -R_{+}^{(s-1)} \x =\frac{d}{dt}\Sigma(t) \label{condition}
\ee
for some $\Sigma(t)$, then
\be
{\widetilde{Q}}=\int_{-\infty}^{0} dx_1 \; 
(J_- ^{(s+1)}+J_+^{(s+1)} -R_-^{(s-1)}-R_+^{(s-1)} )-\Sigma(t)
\ee
is a conserved charge. Note that \ref{condition} depends on the specific
boundary action we are considering.
In this section we prove the integrability of Neumann ($\pd_1 \nv \x=0$), 
Dirichlet ($\nv\x=\nv_0$ a constant, or 
equivalently $\pd_0 \nv\x=0$), and a mixed boundary condition 
(where some components
of $\nv$ satisfy Neumann and the others Dirichlet).
We also analyze the boundary condition proposed by Corrigan and 
Sheng in \cite{CorrSheng} for the $O(3)$ \nls model.
For the GN model we show that the spin 4 charge discussed in the
previous section, with the boundary condition 
$\psi_+^a\x=\e_a\psi_-^a\x$, $\e_a=\pm 1$, 
provides a conserved charge in the boundary case.

\subsection{Nonlinear $\sigma$ Model}

As we explained, we have to look at the combination \ref{condition} 
of the spin 4 currents at
$x=0$ and verify that it can be written as a total time derivative. In our case
the conservation laws are
\bea
\pd_{+}(T_{--})^2\!\!\!\!&=&\!\!\!\!c_1 \;\pd_+ 
(\pd_{-}^2\nv \cdot \pd_{-}^2\nv)+
c_2\;\pd_-(\pd_+\nv \cdot \pd_-\nv \; \pd_-\nv \cdot \pd_-\nv)+
c_3\;\pd_-(\pd_-^3\nv \cdot \pd_+\nv), \nonu
\pd_{-}(T_{++})^2\!\!\!\!&=&\!\!\!\!c_1 \;\pd_- 
(\pd_{+}^2\nv \cdot \pd_{+}^2\nv)+
c_2\;\pd_+(\pd_-\nv \cdot \pd_+\nv \; \pd_+\nv \cdot \pd_+\nv)+
c_3\;\pd_+(\pd_+^3\nv \cdot \pd_-\nv).
\eea
The condition we have to analyze is
\bea
&&(\pd_-\nv\cdot\pd_-\nv)^2 - c_1\;\pd_{-}^2\nv \cdot \pd_{-}^2\nv-
(\pd_+\nv\cdot\pd_+\nv)^2 + c_1\;\pd_{+}^2\nv \cdot \pd_{+}^2\nv- \nonu
&&-c_2\;(\pd_-\nv \cdot \pd_+\nv\; \pd_-\nv\cdot\pd_-\nv) -
c_3\;\pd_-^3\nv \cdot \pd_+\nv + \nonu
&&+c_2\;(\pd_+\nv \cdot \pd_-\nv\; \pd_+\nv\cdot\pd_+\nv) +
c_3\;\pd_+^3\nv \cdot \pd_-\nv\x=\frac{d}{dt}\Sigma(t) \ . \
\label{bcurrent}
\eea
Let us look at the Neumann boundary condition first \footnote{Notice that
we are always considering fields and their derivatives 
at $x=0$ now.}. Since we have
$\pd_1\nv=0$
whenever there is a term like $\pd_\pm \nv$
we can substitute it by $\pd_0 \nv$. The term $\pd_\pm^2 \nv$ becomes
$\pd_0^2\nv\pm\pd_1^2\nv$. 
By appropriately combining
terms in \ref{bcurrent}, we see immediately that they all add up to zero and
we can pick $\Sigma(t)=0$. This means that the Neumann boundary condition
is integrable.

We can now look at the ``dual'' condition to Neumann, namely the Dirichlet
boundary condition, $\nv\x=\nv_0$, constant, 
which is equivalent to $\pd_0\nv\x=0$.
Notice that in this case $\pd_0^n \nv=0$ for any integer $n$.
The manipulations are very similar as in the Neumann case. Wherever there 
is $\pd_{\pm}\nv$
we should replace by $\pm\pd_1\nv$, $\pd_\pm^2\nv$ should be replaced by 
$\pm 2 \pd_0\pd_1\nv+\pd_1^2\nv$, and by using the equations of motion and
the constraint $\nv \cdot \nv=1$, we see that
$\pd_+^3\nv\cdot\pd_-\nv-\pd_-^3\nv\cdot\pd_+\nv=0$. So once again, by 
appropriately collecting terms we see that \ref{bcurrent} vanishes and
we can pick $\Sigma(t)=0$.

Finally we can look at the more general boundary condition, which is 
a mixture of Neumann and Dirichlet in the following sense: take 
$\pd_0 n_i\x=0$ for some collection of indices $\{i\}$, with, 
say, $k$ elements
and $\pd_1 n_j\x=0$
for the remaining $N-k$ indices. Neumann condition is obtained when 
$k=0$, and Dirichlet when $k=N$. The analysis is very similar to the
preceding cases and we shall skip technical comments. The final
conclusion is that this boundary condition too is integrable.

In \cite{CorrSheng} Corrigan and Sheng showed that (classically) the $O(3)$
\nls model on the half-line is integrable if
\be
\pd_1 \nv = -(\kv \times \pd_0 \nv) + 
(\nv \cdot (\kv \times \pd_0 \nv)) \nv \qquad
{\rm and} \qquad \kv\cdot\pd_0\nv=0 \ , \ \label{corsh}
\ee
at $x=0$, $\kv$ arbitrary. By using the equation of motion plus the constraint
$\nv \cdot \nv =1$, this condition is compatible with our spin $4$ current, 
with 
$\Sigma(t)=16 \;c_3\; \pd_0 \nv \cdot \pd_0\pd_1 \nv$. 
This indicates that \ref{corsh} is integrable at the quantum level.

\subsection{Gross-Neveu Model}

Let us consider now the following boundary condition
\be
\psi_+^a\x=\e_a\psi_-^a\x \ , \ \label{bcgn}
\ee
with $a=1,\ldots,N$ and $\e_a=\pm 1$. Before we continue our analysis, we 
should add a few remarks about these conditions. The boundary conditions
\ref{bcgn} can be obtained from the boundary action
\be
{\cal S}_b=\int_{-\infty}^{+\infty} dx_0 \; 
\sum_{a=1}^N \frac{i}{2}\e_a\psi_+^a\psi_-^a
\ . \
\ee
This is the most general form for the boundary 
potential without introducing new
parameters in the theory. 
If we have $N_+$ $\e$'s equal $+1$ and the remaining
$N_-=N-N_+$ $\e$'s equal $-1$,
then we are breaking the original $O(N)$ symmetry at the boundary to
$O(N_+)$ and $O(N_-)$ symmetric sectors. Therefore there are always
two different ways to break the symmetry at the boundary to the same 
groups (pick $N_+$ `$+$' and $N_-$ `$-$', or $N_-$ `$+$' 
and $N_+$ `$-$'),
which will correspond to different CDD factors in the reflection matrices.

Suppose $\e_a$ is different from $\pm 1$ for some $a$. Then we get that
both $\psi_+^a$ and $\psi_-^a$ vanish, which implies that the $a$th fermion
does not propagate, since we have a first order equation of motion. 

Let us now return to the main discussion. Condition \ref{bcgn} 
implies that the 
equations of motion \ref{eomgn} 
give the supplementary condition at the boundary
\be
\pd_-\psi_+^a\x=\pd_+\psi_-^a\x=0 \ , \ \label{eom2}
\ee
since for fermion fields $\psi^2=0$. The boundary condition \ref{bcgn}
can be used along with \ref{eom2} to show that
\be
\pd_0\psi_+^a\x=\pd_1\psi_+^a\x=\e_a\pd_0\psi_-^a\x=-\e_a\pd_1\psi_-^a\x=0
\ . \ 
\ee
We should proceed similarly to the \nls model case, and write down the
correspondent condition from \ref{condition}. There are many more terms
now and the procedure is a bit tedious, but nonetheless, all appropriately
collected terms cancel and we have that we can pick $\Sigma(t)=0$ again.
This shows that the boundary condition \ref{bcgn} preserves integrability at
the quantum level.

\section{Conclusions}

We were able to prove the quantum integrability of the Neumann, 
Dirichlet, and mixed boundary
conditions for the \nls model, and of $\psi_+^a=\pm \psi_-^a$ for the GN model.
The reflection matrices for the \nls model for these conditions
were proposed by Ghoshal in \cite{Ghoshal}. It would be interesting to
investigate the boundary Yang-Baxter equation (BYBE) for this model 
more thoroughly
and see if it is possible to find more general solutions \cite{MacKay}. 
Our results seem
to indicate so, since we have a variety of other boundary conditions for 
the \nls and GN models. 

In \cite{InKoZh} Inami, Konno and Zhang studied, via bosonization, some 
fermionic models on the half-line. In particular, they studied the $O(3)$
GN model and concluded that there were some possible integrable boundary 
conditions of the same form as proposed here \footnote{In \cite{InKoZh}
these authors considered a boundary term of the form $g\psi_+\psi_-$ with
$g \neq \pm i/2$, which should not be included.}. In particular, in the
$O(3)$ GN model it is easy to see that the boundary condition \ref{bcgn}
either preserves full $O(3)$ invariance or it breaks it to $O(2)$. In
each case there are two possibilities, in a similar fashion to the boundary
Ising model \cite{gzam}. The difference between the two correspondent 
reflection matrices will appear as CDD prefactors.

Another connection we can make to results in the literature is the following.
It is possible to relate the $O(2N)$ GN model to the 
affine Toda field theories (ATFT) with imaginary coupling, 
using bosonization \cite{Shankar,MorDeMart}, in a similar 
fashion to the way Witten used to establish the mapping from the $O(4)$
GN model to two decoupled sine-Gordon models. The ATFT (with
real coupling) on the
half-line were considered by Bowcock, Corrigan, Dorey and Rietdjik 
in \cite{BoCoDoRi}, where they found that there is only a discrete set of 
integrable boundary conditions. It would be interesting to investigate
the relation between their results and our boundary conditions.

One interesting direction to pursue would be to study the most general
integrable boundary conditions for these models, compatible with the spin 3
charge that we have analyzed.

As a last remark, it should be interesting to apply our considerations to the
local charges in the principal chiral model studied by Evans, Hassan and 
Mackay in \cite{EvHaMac}.

\section*{Acknowledgments}
We would lik to thank K. Schoutens for a critical
reading of the manuscript. ADM would like to thank
useful discussions with F. Morales and C.A. Scrucca. 
MM would like to thank S.F. Hassan and
G. Thompson for useful conversations.

\newpage
\appendix
\section{Appendix}

As we mentioned in section 2, here we list the possible anomaly terms that
can appear in the right hand side of $\pd_-(T_{++})^2=0$ \footnote{There is
an analogous analysis for the $\pd_- (T_{++})^2=0$ conservation law.}. 
These operators have to have Lorentz weight $3$, 
dimension $5$, and by using equations of motion its easy to show that
we can restrict ourselves to operators that are
zeroth order in $\pd_-$ and quadratic in $\psi_-^a$. Witten has shown in
\cite{Witten} that all such operators can be written as $\pd_\pm$ of something.
We are actually interested in the {\em something} structure of these operators.
This means that we have to look for operators that have dimension $4$ 
and Lorentz
weight $4$ (from $\pd_-$ of something) or operators with dimension 4 and
Lorentz weight $2$ (from $\pd_+$ of something). The list is as follows
\begin{enumerate}

\item{\em Dimension $4$ and Lorentz weight $4$}
\bea
&&\psi_+^a \partial_+^3 \psi_+^a \,\, ,\nonu
&&\partial_+\psi_+^a \partial_+^2 \psi_+^a \,\, ,\nonu
&&\psi_+^a \partial_+ \psi_+^a \, \psi_+^b \partial_+ \psi_+^b \,\, \nn .
\eea

\item{\em Dimension $4$ and Lorentz weight $2$}
\bea
&&\psi_-^a \partial_+^3 \psi_-^a \,\, ,\nonu
&&\partial_+\psi_-^a \partial_+^2 \psi_-^a \,\, ,\nonu
&&\psi_-^a \partial_+^2 \psi_+^a \, \psi_+^b \psi_-^b \,\, ,\nonu
&&\psi_-^a \partial_+ \psi_+^a \, \psi_-^b \partial_+ \psi_+^b \,\, ,\nonu
&&\psi_+^a \partial_+ \psi_+^a \, \psi_-^b \partial_+ \psi_-^b \,\, ,\nonu
&&\partial_+ \psi_+^a \partial_+ \psi_-^a \, \psi_+^b \psi_-^b \,\, ,\nonu
&&\psi_+^a \partial_+^2 \psi_+^a \, \psi_+^b \psi_-^b \,\, ,\nonu
&&\psi_+^a \partial_+ \psi_+^a \, (\psi_+^b \psi_-^b)^2 \,\, \nn.
\eea

\end{enumerate}

\end{document}